\documentstyle [aps,twocolumn]{revtex}

\begin{document}

\title{Statistics of pre-localized states in disordered conductors}

\author{Vladimir I.  Fal'ko $^{1,3}$ and K.B.  Efetov$^{2,4}$}

\address{$^1$ Max-Planck-Institut f\"ur Fesk\"orperforschung, Heisenbergstr.
1, 70569
Stuttgart, Germany \\
$^2$ Max-Planck-Institut f\"ur Physik Komplexer Systeme, Heisenbergstr. 1,
70569 Stuttgart, Germany \\
$^3$ Institute of Solid State Physics RAS, Chernogolovka, 142432 Russia \\
$^4$ L.D.  Landau Institute for Theoretical Physics, Moscow, Russia}

\date{\today{}}

\maketitle

\begin{abstract}

The distribution function of local amplitudes, $t=|\psi ({\bf r}_o
)|^2$, of
single-particle states in disordered conductors is calculated exactly on
the basis of the supersymmetric $\sigma$-model approach using a
saddle-point solution of its reduced version.  Although the
distribution of relatively small amplitudes can be approximated by
the universal Porter-Thomas formulae known from the random matrix
theory, the statistics of large $t$'s is strongly modified by localization
effects.  In particular, we find a multifractal behavior of eigenstates
in 2D conductors which follows from the non-integer power-law
scaling for the inverse participation numbers (IPN) with the size of
the system, $Vt_n\propto L^{-(n-1)d^{*}(n)}$, where $d^{*}(n)=2-\beta^{
-1}n/(4\pi^2\nu D)$ is a
function of the index $n$ and disorder.  The result is valid for all
fundamental symmetry classes (unitary, $\beta_u=1$; orthogonal, $\beta_
o={1\over 2}$, and
symplectic, $\beta_s=2$).  The multifractality is due to the existence of
pre-localized states which are characterized by power-law envelopes
of wave functions, $|\psi_t(r)|^2\propto r^{-2\mu}$, $\mu =\mu (t)
<1$.  The pre-localized
states in short quasi-1D wires have the power-law tails $|\psi (x)
|^2\propto x^{-2}$,
too, although their IPN's indicate no fractal behavior.  The
distribution function of the largest-amplitude fluctuations of wave
functions in 2D and 3D conductors has logarithmically-normal
asymptotics.

\end{abstract}

\pacs{73.20.Dx,71.25.-s,05.45.b}

\newpage

\section{Introduction}

Localization of a particle by a random potential has been extensively
investigated during the past several decades
\cite{Anderson,Thouless,Mott,Wegner1,WeakLoc,KramerMcKinnon}.  It is
well known \cite{WeakLoc,KramerMcKinnon,Efetov} that, at strong
disorder, single-particle wave functions are confined and have
exponentially decaying tails beyond the scale of the localization length
$L_c$.  At weak disorder, the localization length can be very large in 1D
and 2D conductors, and infinite in 3D.  A natural question arises:
What is the behavior of the wave functions at distances smaller than
the localization length?  Despite of its importance, the problem of
structure of quantum states of weakly disordered conductors for
scales below the length $L_c$ has only recently started to attract
interest
\cite{Wegner,Alt,Aoki,Schreiber,Kramer,Ohtsuki,Evangelou,Jan,Cast,Dima}.

In particular, one of the issues that has not been explored up to now
concerns the way the localized states develop as a consequence of
the increase of disorder in an isolated piece of a metal, though a
great deal is already known about the extended states in it.  Some
part of recent results related to extended (metallic-type) states have
been obtained by mapping the problem of quantum mechanics in the
classically chaotic systems to the Wigner-Dyson random matrix
theory \cite{Gorkov,Brody,Berry,Stone}, or using the zero-dimensional
supermatrix $\sigma$-model \cite{Efetov,EP,Simons,Prigodin,FalkoEf,FM},
which are two equivalent ways of describing disordered and chaotic
systems.

Both the advantage and disadvantage of such an approach comes from
the statistical equivalence of eigenstates which is usually built into
the construction of the random matrix substituting the real dynamics.
This reveals the set of universalities of the spectra, the level-level
correlations and the transition matrix elements \cite{Brody} which
are similar for a wide variety of objects.  For example, the
distribution function of local densities of wave functions $|\psi
({\bf r}_o)|^2$ in a
chaotic cavity which one can find in such a way is determined only
by the fundamental symmetry of the system and its volume $V\sim L^
d$,
but is independent of the level of disorder (i.e.,  of the value of a
mean free path $l$) or a physical dimension, $d$.

On the other hand, this approach hides individual features of
physically different systems and permits to describe {\it only
metallic-type states\/} which equally test the random potential all over
the sample.  More complex states which can distinguish between the
ballistic and diffusive regimes have to be analyzed beyond the
conventional random matrix theory.  Numerical evidence for their
existence have been obtained by several groups
\cite{Aoki,Schreiber,Kramer,Jan}.  The goal of the theory to be
presented in the present paper is to find manifestations of these
{\it precursors of localization\/} among the wave functions of classically
diffusive conductors ($p_Fl\gg 1$, $l\ll L$).  That is, we consider an isolated
piece of a disordered metal with dimensions $l\ll L<L_c$, assuming that
the internal 'conductance' $g$ which one would assign to the 'electric
circuit' connecting the observation point (blown up to the mean free
path size) with the external surface of the specimen \cite{Rem1} is
much larger than the conductance quantum, i.e.  $g\gg 1$.  In the
following Sections, we perform a statistical analysis of local
densities and partly reconstruct spatial structure of those rare
states which have locally too high amplitudes (as compared to the
average $V^{-1}$) to fit into the universal random matrix theory
description.  As can be suggested on the basis of the theory below,
these states are responsible for non-Gaussian tails of distributions of
fluctuations of local densities of states and conductances suggested
by Altshuler, Kravtsov and Lerner \cite{Alt} and are generic for the
long-living current relaxation discussed in Refs.  \cite{Alt,Dima}.

Our paper is organized as follows.  In Section II, we introduce the
notion of the eigenstates statistics (II.A), discuss the universal
distributions of metallic-type states (II.B) and, then, sketch the main
results of the paper focusing our attention at the localization effects
(II.C).  Sections III and IV are devoted to the presentation of our
theoretical scheme:  We derive a reduced supersymmetric $\sigma$-model
and show that it has a non-trivial saddle-point.  The details of the
derivation of the saddle-point solutions of the reduced $\sigma$-model are
given separately for each of the fundamental symmetry classes
(unitary, orthogonal and symplectic) in Sections IV.A-C, and the
influence of fluctuations around the saddle-point is discussed in
Section IV.D and the Appendices.  The resulting statistics of wave
functions and the structure of the pre-localized states in the
conducting regime in 1D, 3D and in the most interesting case of 2D
samples are discussed in Sections V, VI and VII, respectively.
Section VIII contains a brief summary of our results and their
discussion.

\section{Metallic versus pre-localized states in the eigenstates
statistics (preliminaries and results)}

In this Section, we give mathematical formulation to the problem of
the eigenvalues statistics in disordered systems and consider
alternative approaches to its solution.  That is what is the
subsection A about.  The next part B is devoted to the universal
statistics of metallic-type of states known in the random matrix
theory as the Porter-Thomas distribution.  The localization effects
which are beyond the random matrix theory approach are discussed in
the subsection C, where we give an essence of the obtained results.
This subsection is written for the first reading and can be used as a
guide through the rest of the text.

\subsection{Definitions of the eigenstates statistics}

To define the statistics which we shall be studying in this paper, we
first mention that the properly normalized eigenstates $\{\psi_{\alpha}
\}$ we
consider below correspond to a quantum particle in a disordered
cavity
$$\left\{{{{\bf P}^2}\over {2M}}+U({\bf r})\right\}\psi_{\alpha}({\bf r}
)=\epsilon_{\alpha}\psi_{\alpha}({\bf r}),\;\psi_{\alpha}({\bf r}\in
S)=0,$$
where $U$ is a random potential.  The local amplitude $\psi$ of a wave
function at some observation point ${\bf r}_o$ inside the sample, i.e.,

\begin{equation}\label{IA1}
$$t\equiv |\psi ({\bf r}_o)|^2,$$
\end{equation}
will be the object of our statistical analysis.  In this we employ
two related quantities:  the distribution
function $f(t)$ of local amplitudes $t$ averaged over disorder,

\begin{equation}\label{IA2}
$$f(t)=\Delta\left\langle \sum_{\alpha}\delta\left(t-|\psi_{\alpha}
({\bf r}_o)|^2\right)\delta (\epsilon -\epsilon_{\alpha})\right\rangle
,$$
\end{equation}
and the set of generalized inverse participation numbers (IPN)
\cite{Wegner,Jan} which are the moments of the distribution function
$f$,

\begin{equation}\label{IA3}
$$t_n=\Delta\left\langle \sum_{\alpha}|\psi_{\alpha}({\bf r}_o)|^{
2n}\delta (\epsilon -\epsilon_{\alpha})\right\rangle \equiv\int_0^{
\infty}t^nf(t)dt.$$
\end{equation}
As indicated, $\left<\right>$ denotes the averaging over random configurations
of a
random potential $U$ in the
system. In Eq. (\ref{IA3}), the sum is over the full set of states $
\{\psi_{\alpha}\}$,
$V$ is the volume of the system, and $\Delta =$ $\left(\nu V\right
)^{-1}$ denotes the mean level
spacing with $\nu =\nu (\epsilon )$ the density of states per unit volume.
Since
the distribution function $f$ and wave functions $\{\psi_{\alpha}\}$ are
normalized,
one has the following relations:
$$t_0=\int_0^{\infty}f(t)dt\equiv 1;\;t_1=\left<|\psi_{\alpha}|^2\right
>\equiv V^{-1}.$$
One can also introduce the distribution $f_s(\sigma ,t)$ of a local
spin-density
of the wave with $\sigma =\downarrow ,\uparrow$.  It is an important quantity
for systems
with a strong spin-flip scattering.  In random matrix theories these
are known as a symplectic ensemble.  In such a case, the statistics
can be formulated in terms of spin-projected eigenstates, e.g.
$t_{\downarrow}=|\psi_{\downarrow}({\bf r}_o)|^2$ \cite{Rem2}.  The
distribution of a total local density
($t=t_{\downarrow}+t_{\uparrow}$) can be found as the convolution

\begin{equation}\label{IA4}
$$f_s(t)=\int_0^tf(\downarrow ,t-t^{\prime})f(\downarrow ,t^{\prime}
)dt^{\prime}.$$
\end{equation}

Historically, the studies of eigenstates in disordered conductors
started from Wegner's perturbative calculations of IPN's
\cite{Wegner}.  Due to the equivalence between the descriptions based
on the distribution function and the full set of its moments
\cite{Math}, in most of the later studies \cite{Alt,EP,MF} the
eigenstates statistics were reconstructed from the set of IPN's.
Alternatively, one can start from calculating directly the entire
distribution function \cite{EP,MF1,FalkoEf,Multi}, especially regarding
the possibility to apply the supersymmetry technique \cite{Efetov}.
This alternative approach has already been used for describing the
eigenstates statistics over the entire crossover regime from the
orthogonal to unitary ensembles (low magnetic fields) \cite{FalkoEf}.
More recently, this construction has been advanced by developing a
reduced $\sigma$-model which is applicable to closed systems.  The reduced
$\sigma$-model has non-trivial saddle-point solutions which enabled us to
consider the localization effects non-perturbatively \cite{Multi}.

The idea to work with the distribution function as a whole has also
the additional advantage that it makes possible to select those rare
states which do not fit to the universal statistics and study their
spatial structure.  The latter information is implicit in the
cross-correlation function $R(t,r)$,

\begin{equation}\label{IA5}
$$R(t,r)=\Delta\left\langle \sum_{\alpha}\delta\left(t-|\psi_{\alpha}
({\bf r}_o)|^2\right)|\psi_{\alpha}({\bf r}_o+{\bf r})|^2\delta (\epsilon
-\epsilon_{\alpha})\right\rangle .$$
\end{equation}
As it will be clear from the calculations below, rare pre-localized
states show up as deviations of the function in Eq.  (\ref{IA2})
from universal distributions at the tails where $t\gg V^{-1}g$, so that the
combination $R(t,r)/f(t)$ mimics the envelope $|\psi_t(r)|^2$ of these
states at distances $r=|{\bf r}|>l$ from the top-amplitude ($t$) position.

\subsection{Metallic states and universal statistics}

To find the manifestation of the pre-localized states in the
distribution function $f(t)$, we have, for comparison, to give an idea
about what would be the form of the distribution function if all
states were extended.  Qualitatively, the extended states test the
realization of a random potential equivalently all over the sample and
that's why their statistics coincides with the Porter-Thomas
eigenstates statistics renown in the random matrix theories
\cite{Berry,Brody}.  The recent studies of the properties of the
eigenstate of disordered and ballistic chaotic cavities (both using the
numerical tools \cite{Berry,Stone} or based on the zero-dimensional limit
(0D) of the supersymmetric non-linear $\sigma$-model
\cite{Efetov,EP,Simons,FM}) have confirmed such an expectation.

Depending on the fundamental symmetry class, the Porter-Thomas
distributions can be represented as following.  For the single-particle
Hamiltonian describing a spin-less particle in the system with a
broken (e.g.,  by a magnetic field) time-reversal symmetry (unitary
class), the distribution function of local amplitudes and the
corresponding IPN's have the form

\begin{equation}\label{IB1}
$$f_u(t)=V\exp\left\{-Vt\right\},\;t_n=n!V^{-n},$$
\end{equation}
whereas in the case of a system with the time-reversal symmetry
(orthogonal ensemble),

\begin{equation}\label{IB2}
$$f_o(t)=\sqrt {{V\over {2\pi t}}}\exp\left\{-{{Vt}\over 2}\right\}
,\;t_n={{(2n-1)!!}\over {V^n}}.$$
\end{equation}
For spin-${1\over 2}$ particles which undergo a strong spin-orbit interaction
(symplectic ensemble) the Porter-Thomas-type of a distribution can be
repeated both for the spin-projected wave functions and for the total
density and has the form

\begin{equation}\label{IB3}
$$f(\downarrow ,t)=2Ve^{-2Vt};\;f_s(t)=4V^2te^{-2Vt}.$$
\end{equation}

\subsection{Localization effects and eigenstates statistics beyond the
universal limit}

The universal statistics described by Eqs.  (\ref{IB1}-\ref{IB3}) are
presented only as a reference point for the subsequent analysis.  The
rare events which cannot be described on the basis of the statistical
equivalence of eigenstates have to be studied using more
sophisticated methods, and need the extension of the non-linear
$\sigma$-model beyond the 0D limit.  Details of these calculation are
presented in Sections IV-VII, whereas in the forthcoming subsection
we sketch only the basic results.  In few words, the universal
disorder-independent laws work well enough either until this
disorder is so weak, that the system behaves as in the nearly
ballistic regime, or at small amplitudes $t<V^{-1}\sqrt g$.  But they are
partly broken or, at least, modified after the disorder makes the
electron motion diffusive.  In one- and two- dimensional conductors,
this requires a different statistical treatment of states which have
too high splashes of a local amplitude, $t>V^{-1}\sqrt g$.

The method of taking into account all finite (i.e.,  not only small)
inhomogeneous variations of the fields used in the supersymmetric
field theory is based on the existence of saddle-point solutions of the
non-linear $\sigma$-model discovered by Muzykantskii and Khmelnitskii
\cite{Dima}.  The existence of the saddle-point solution is especially
prominent for a reduced version of $\sigma$-model formulated and solved in
Ref.  \cite{Multi}.  An interesting result of Ref.  \cite{Multi} for the
{\it quantum diffusion in the dimension $d=2$\/} is the {\it multifractality\/}
of
the states which is in agreement with previous numerical simulations
\cite{Schreiber,Jan}.  Using the same saddle-point method as for the
unitary ($u$) symmetry class \cite{Multi}, we extend the analysis of 2D
systems to the other fundamental symmetry classes (orthogonal, $o$;
symplectic, $s$; spin-unitary \cite{Spin-unitary}, $su$) and arrive, again,
at the multifractality.  The latter is manifested by the following
scaling of INP's,

\begin{equation}\label{IC1}
$$Vt_n\propto L^{-(n-1)d^{*}},\;d^{*}(n)=2-{{\beta^{-1}n}\over {4\pi^
2\nu D}},\quad\matrix{\beta_u=1\cr
\beta_o={1\over 2}\cr
\beta_{s,su}=2\cr}
.$$
\end{equation}
The fractal (or generalized R\'enyi \cite{Mandelbrot}) dimensions,
$d^{*}(n)$, obey Eq.  (\ref{IC1}) only for those $n$'s where they are positive
and are obtained in the leading order in
$(2\pi\nu D)^{-1}$, where $D$ is the classical diffusion coefficient.
The sensitivity of the derived statistics to boundary conditions, as
well as the form of the correlation function $R(t,r)$
which we find in our calculations, enables us to suggest such a
behavior of 2D multifractal states which is associated with the
power-law envelopes of their tails, $|\psi ({\bf r}_o+{\bf r})|^2\propto
(l/r)^{2\mu}$.  Being
extended from the position of a rare high amplitude splash
$|\psi ({\bf r}_o)|^2=t\gg 1$, these tails have exponents $\mu (t)$ individual
for each
state marked by its own $t$.

The behavior of {\it pre-localized states in a quasi-1D wire\/} within the
localization length scale $L_c$ also resembles the power-law localization.
Independently of $t$, $|\psi (x_o+x)|^2\propto L_cx^{-2}$.  The density of wave
functions accumulated by these tails is integrable, so that no
assertion about fractality can be made and the inverse participation
numbers $t_n$ tend to take a volume-independent form for $n>g$.

The {\it localization effects in 3D conductors\/} are weak, if disorder is
weak enough to keep the system far away from the metal-insulator
transition, $p_Fl\gg 1$.  As a result, the statistics of eigenstates in 3D
conductor is most similar to the universal one:  Amplitude $t$ scales
with $V$, and the mean free path $l$ appears only as an extra
parameter both in the distribution function  $f=f(Vt,p_Fl)$, and inverse
participation numbers $t_n\propto V^{-n}\kappa (n,p_Fl)$.

Nevertheless, even then, the statistics of rare events shows
intriguing deviations from the Porter-Thomas formulae.  Both for 2D
and 3D diffusive samples we obtain the {\it logarithmically-normal
distribution\/} of large local amplitudes of wave functions,

\begin{equation}\label{IC2}
$$f(t)\approx\exp\left\{-\beta{{\pi^2\nu\!D}\over {\eta_d}}\ln^2T\right
\},\quad T={{Vt\eta_d}\over {2\pi^2\nu D}},$$
\end{equation}
where $\eta_2=\ln{L\over l}$  and $\eta_3\sim (2l)^{-1}$.
Although we study an isolated specimen, this result strikingly
resembles the asymptotics of distributions of local density of states
or conductance fluctuations found in open systems \cite{Alt}.  This
signals about deep physical reasons behind it related, probably, to the
properties of random walk paths.

\section{Eigenstates problem in terms of a non-linear supersymmetric
$\sigma$-model}

In the following paragraphs, we formulate the eigenstates statistics
problem in terms of the non-linear $\sigma$-model.  The details of this
technique are described in the review article \cite{Efetov}, and below
we give only a compressed extraction from it, keeping similar
notations.

One can try to use the supersymmetry technique as soon as a
physical quantity of interest is expressed in terms of retarded and
advanced Green's functions,

\begin{equation}\label{II1}
$$G_{\epsilon}^{R,A}({\bf r},{\bf r})=\sum_{\alpha}{{|\psi_{\alpha}
({\bf r})|^2}\over {\epsilon -\epsilon_{\alpha}\pm i\gamma /2}}.$$
\end{equation}
In Eq. (\ref{II1}), $\gamma$ is a level broadening.  In an isolated sample, one
has to take
the limit of $\gamma\to 0$.  Due to the discreteness of the spectrum of levels
$\{\epsilon_{\alpha}\}$, this extracts only $\epsilon_{\alpha}$ the closest to
the current energy slice
$\epsilon$.  Using the expression in Eq.  (\ref{II1}) and taking the limit of $
\gamma\to 0$,
one can so formalize the statistics of Eq.  (\ref{IA2}) that
\cite{FalkoEf}

$$f(t)=\Delta\lim_{\gamma\to 0}\left<\int{{d{\bf r}^{\prime}}\over {~
2\pi i}}\left(G_{\epsilon}^A({\bf r}^{\prime},{\bf r}^{\prime})-G_{
\epsilon}^R({\bf r}^{\prime},{\bf r}^{\prime})\right)\right.\quad$$
\begin{equation}\label{II2}
$$\left.\qquad\delta\left(t-i\gamma VG_{\epsilon}^R({\bf r},{\bf r}
)\right)\right>.$$
\end{equation}

The reformulation of Eq.  (\ref{II2}) in terms of the $\sigma$-model exploits
the fact \cite{EP} that any product of Green functions,

\begin{equation}\label{II3}
$$i^{n+1}\left[G_{\epsilon}^R({\bf r},{\bf r})\right]^nG_{\epsilon}^
A({\bf r}^{\prime},{\bf r}^{\prime})=\qquad$$
\end{equation}
$$\qquad ={{-1}\over {n!}}\int |s_1({\bf r}^{\prime})|^2|s_2({\bf r}
)|^{2n}e^{-L(\Psi )}D\Psi ,$$
can be represented as a functional integral over the 8-component
super-vector field $\Psi =\left(\Psi_1,\Psi_2\right)$, $\Psi_m={1\over {\sqrt
2}}\left(\chi^{*}_m,\chi_m,s^{*}_m,s_m\right)$.  The
super-vector $\Psi$ is composed of 4 anti-commuting and 4 commuting
variables $\chi$ and $s$, respectively.  The indices $m=1,2$ appear in order
to distinguish between advanced and retarded Green functions.
Besides $\Psi$, the charge-conjugate field $\bar{\Psi}$ should be defined; one
can find
this definition in Ref.  \cite{Efetov}.  The action $L(\Psi )$,

\begin{equation}\label{II4}
$$L(\Psi )=i\int\bar{\Psi }({\bf r})[\epsilon -\hat {H}_0-U({\bf r}
)-i{{\gamma}\over 2}\Lambda ]\Psi ({\bf r})d{\bf r},$$
\end{equation}
$$\Lambda^{11}=-\Lambda^{22}=\hat {1}$$
incorporates both the free-particle Hamiltonian $\hat {H}_0$ and the random
impurity potential $U({\bf r})$.

After Gaussian averaging over $U$ \cite{Efetov}, we derive a new
Lagrangian with an interaction of the super-fields $\Psi$.  The interaction
term can be decoupled by the functional integration over a
super-matrix field $Q$, so that any calculation is finally reduced to
the evaluation of a functional integral,

\begin{equation}\label{II5}
$$\int DQ\exp \{-F[Q]\}W(Q),$$
\end{equation}
over the slow-varying superfields $Q({\bf r})$.  This manipulation is
analogous to the introduction of an effective order parameter in the
theory of superconductivity.  The free energy which determines
weights of different configurations of $Q$ appears after integrating over
fast modes and has the form

\begin{equation}\label{II6}
$$F[Q]=\int d{\bf r}\left[-{1\over 2}{\rm S}{\rm t}{\rm r}\ln\left
(-i\hat H_0+{{\gamma}\over 2}\Lambda +{Q\over {2\tau}}\right)+{{\pi
\nu}\over {8\tau}}{\rm S}{\rm t}{\rm r}Q^2\right].$$
\end{equation}
The 'anomalous mean' $Q$ can be found from the self-consistency
condition

\begin{equation}\label{II6a}
$$\pi\nu Q=\int dp\left(-i\hat H_0+\gamma\Lambda /2+Q/2\tau\right)^{
-1}$$
\end{equation}
which minimizes $F[Q]$.

In the limit of $\gamma\to 0$, the solutions of Eq. (\ref{II6a}) take the
values

\begin{equation}\label{II7}
$$Q=V\Lambda\bar {V}$$
\end{equation}
from the degeneracy space of one of the graded symmetry group
\cite{DeWitt}.  This field-theoretical model is strongly non-linear,
since the matrix $V$ satisfies the condition $\bar {V}V=1$,  and operations of
the conjugation $V\to\bar {V}$ and the super-trace (Str) in Eq.
(\ref{II6}) are those defined in Ref.  \cite{Efetov}.  The 'rotation' $
V$,

\begin{equation}\label{II8}
$$V=\left(\matrix{u_1&0\cr
0&v_1\cr}
\right)\exp\left(\matrix{0&-iu_2{{\hat{\theta}}\over 2}\bar {v}_2\cr
-iv_2{{\hat{\theta}}\over 2}\bar {u}_2&0\cr}
\right),$$
\end{equation}
is parametrized with the equal number of commuting and
anti-commuting variables.  The second matrix in the product in Eq.
(\ref{II8}) is composed only of commuting ones.  All anti-commuting
variables are collected into the matrices $u_1$ and $v_1$.  Smooth spatial
variations of the $Q$-field at the length scale longer than the mean free
path $l$ and the influence of a small but finite value of $\gamma$ can be taken
into account by the effective free energy functional,

\begin{equation}\label{II9}
$$F[Q]={{\pi\nu}\over 4}\int d{\bf r}{\rm S}{\rm t}{\rm r}\left[{D\over
2}\left(\nabla Q\right)^2-\gamma\Lambda Q\right].$$
\end{equation}
The extension of this equation to the case of spin-${1\over 2}$ particles can
be found in Ref.  \cite{Efetov}, too.

To transform the formulae in Eq.  (\ref{IA2}-\ref{IA5}) into integrals
over the $Q$-space, we expand the $\delta$-function in Eq.  (\ref{II2}) into
the
formal series in $G_{\epsilon}^R$, and study the averages for all $
n$
$$\left\langle i^{n+1}\left[G_{\epsilon}^R({\bf r},{\bf r})\right]^
n\int d{\bf r}^{\prime}G_{\epsilon}^A({\bf r}^{\prime},{\bf r}^{\prime}
)\right\rangle .$$
Using Eq.  (\ref{II3}), each of them can be represented as a functional
integral over the field $\Psi$ and then modified into the construction
$$\lim_{\mu ,\lambda\to 0}\int{{d\zeta_1d\zeta_2}\over {(2\pi )^2}}{{
(-1)^nn!}\over {2(2n)!}}\partial_{\mu}\partial_{\lambda}^n\left\langle
\int D\Psi e^{-L(\Psi )-\delta L(\Psi )}\right\rangle .$$
In the latter equation, we perform an additional integration over the
phases $\zeta_{1,2}$ (which are hidden into the vectors
$\bar {v}_1=\sqrt 2(0,0,e^{i\zeta_1},e^{-i\zeta_1},0,0,0,0)$, $\bar {
v}_2=\sqrt 2(0,0,0,0,0,0,e^{i\zeta_2},e^{-i\zeta_2})$ and add to
the Lagrangian from Eq.  (\ref{II4}) a weak perturbation $\delta L$,
$$\delta L=\int d{\bf r}^{\prime}\left[\mu (\bar v_1\Psi ({\bf r}^{
\prime}))(\bar\Psi ({\bf r}^{\prime})v_1)\right.\qquad$$
$$\qquad\left.+\lambda (\bar v_2\Psi ({\bf r}^{\prime}))(\bar\Psi
({\bf r}^{\prime})v_2)\delta ({\bf r}^{\prime}-{\bf r})\right].$$

After this, we have to evaluate the generating functional
$\left\langle \int D\Psi e^{-L(\Psi )-\delta L(\Psi )}\right\rangle $.  Its
exponent differs from that in Eqs.
(\ref{II5}-\ref{II9}) only by the perturbation
$$\delta\hat {H}=i\int d{\bf r}^{\prime}\left[\mu v_1\otimes\bar v_
1+\lambda v_2\otimes\bar v_2\delta ({\bf r}^{\prime}-{\bf r})\right
]$$
added to the Hamiltonian $\hat {H}_0$ in Eq.  (\ref{II6}).  The latter results
in
an additional term $\delta F$ in the free energy functional; we find $
\delta F$ by
expanding the logarithmical expression in Eq.  (\ref{II6}) into the
series in $\mu$ and $\lambda .$ Doing that, we keep only the contributions up
to
the first order in $\mu$, whereas 'cross-terms' which originate from
pairing of $\Psi$'s at different coordinates (${\bf r}$ and ${\bf r}^{
\prime}$) can be neglected.  As
an intermediate step, we obtain
$$\delta F={{\mu}\over 2}\pi\nu (\bar {v}_1Qv_1)+{1\over 2}\ln\left
[1+\lambda\pi\nu (\bar v_2Qv_2)\right],$$
and after some algebra we arrive at

\begin{equation}\label{II10}
$$f(t)=\int{{d\zeta}\over {2\pi}}\lim_{\gamma\to 0}\int {\rm D}Q\left
[\int{{d{\bf r}}\over {4V}}{\rm S}{\rm t}{\rm r}(\pi_1Q({\bf r}))\right
]$$
\end{equation}
$$\delta\left(t-{{\pi\nu\gamma}\over 2}{\rm S}{\rm t}{\rm r}(\Upsilon
(\zeta )Q({\bf r}_o))\right)e^{-F[Q]}$$
where
$$\pi_1=\left(\matrix{\pi_b&0\cr
0&0\cr}
\right),\;\pi_2=\left(\matrix{0&0\cr
0&\pi_b\cr}
\right),\;\;\pi_b=\left(\matrix{0&0\cr
0&1\cr}
\right)\otimes\tau_0,$$
and
$$\Upsilon (\zeta )=\pi_2e^{i\zeta\tau_3}(\tau_0+\tau_1)e^{-i\zeta
\tau_3}.$$
Everywhere below, $\tau_i$ are the Pauli matrices, and $\tau_0$ is $
2\times 2$ unit
matrix.

\section{Reduced $\sigma$-model and its saddle-point solutions}

Basing on Eq.  (\ref{II10}), we can obtain the full statistics of local
amplitudes $|\psi |^2$ for any regime.  As we mentioned before, the
universal expressions of Eqs.  (\ref{IB1}-\ref{IB3}) can be rederived by
assuming the zero-dimensional (0D) limit:  the coordinate-dependent
field $Q({\bf r})$ has to be replaced by its value at the observation point,
$Q({\bf r}_o)\equiv Q_o$, which transforms the functional integral in Eq.
(\ref{II10}) into a definite integral over $Q_o$.

To go beyond the 0D approximation, one should take into account
inhomogeneous fluctuations of the field $Q$.  If we integrate over
$Q_o=V_o\Lambda\bar {V}_o$, any of functions in Eqs.  (\ref{IA2}-\ref{IA4}) can
be
finally expressed in terms of relative rotations of the $Q$-field with
respect to its value at ${\bf r}_o$.  This is the {\it reduced
$\sigma$-model\/}.  For its
derivation, it is significant to note that the degeneracy space of the
supermatrix $Q$ is non-compact.  Due to this property, the main
contribution to the integral in Eq.  (\ref{II10}) comes from the region
of large $Q_o$'s where ${\rm S}{\rm t}{\rm r}(\Upsilon Q_o)\propto
1/\gamma$.  As a result, finite variations of
$Q({\bf r})$ produced by means of local rotations $Q({\bf r})$ $\to$
$Q({\bf r}^{\prime})=V({\bf r},{\bf r}^{\prime})Q({\bf r})V({\bf r}
,{\bf r}^{\prime})$ of the supermatrix field along the
non-compact 'direction' can be taken into account consistently, since
they cover only relatively small environs of an 'infinitely large' $
Q_o$.

Using the decomposition $V\left({\bf r}\right)=V_o\tilde {V}\left(
{\bf r}\right)$, we define supermatrices $\tilde {Q}$ of
the reduced $\sigma$-model as

\begin{equation}\label{III1}
$$\tilde {Q}=\tilde {V}\Lambda\tilde{\bar {V}},\quad Q({\bf r})=V_
o\tilde {Q}({\bf r})\bar {V}_o,\quad\tilde {Q}({\bf r}_o)=\Lambda
.$$
\end{equation}
Due to the invariance of the $Q$-space, the transformation of Eq.
(\ref{III1})  preserves the form of the gradient term in the free
energy $F$ in Eq.  (\ref{II9}), whereas the second term in $F$ can be
modified as
$$F_2=-{{\pi\nu\gamma}\over 4}\int d{\bf r}{\rm S}{\rm t}{\rm r}(\tilde {
Q}_o\tilde {Q}\left({\bf r}\right)),\quad\tilde {Q}_o=\bar {V}_o\Lambda
V_o.$$
A corresponding substitution can be done in the pre-exponential in Eq.
(\ref{II10}), too.  The explicit form and exact parametrization of the
matrix $Q_o$ varies from one symmetry ensemble to another.
Nonetheless, in the limit of $\gamma\to 0$, those parameters of the $
Q$-space
which are responsible for its non-compactness appear in the
argument of the $\delta$-function in Eq.  (\ref{II10}) in the same
combination with the factor $\gamma$ that enters to the 'potential' part of
the free energy, $F_2$.  Therefore, integrating over $Q_o$ in this limit, we
eliminate $\gamma$ and convert Eq.  (\ref{II10}) into expressions which
relate the distribution function $f(t)$ to the generating functionals
represented in terms of the fields $\tilde {Q}$.

The parametrization of $Q$-matrices, and, therefore, derivation and
form of the reduced $\sigma$-model depend on the fundamental symmetry of
the system.  In parts A, B and C of this Section we specify this for
unitary, orthogonal and symplectic symmetry classes separately,
though it turns out that the most essential part of our calculation -
the use of the saddle-point solution described in subsection IV.D - is
quite similar for all of them.

\subsection{Unitary ensemble}

In the unitary case, the parametrization of the
$Q$-field using Eq.  (\ref{II8}) includes 'angles'
$$\hat{\theta }=\left(\matrix{\theta\tau_0&0\cr
0&i\theta_1\tau_0\cr}
\right),\quad\matrix{\,0<\theta <\pi\cr
0<\theta_1<\infty\cr}
,$$
where only one of them is imaginary and makes the symmetry group
non-compact.  Matrices $u_2$ and $v_2$ in Eq.  (\ref{II8}) are diagonal and
can be trivially eliminated from Eq.  (\ref{II10}) as well as the
external phases $\zeta$.  Other details of the integration over $Q_
o$ are the
same as those in Ref.  \cite{FalkoEf}.

The distribution function $f$,

\begin{equation}\label{IIIA1}
$$f_u(t)={1\over V}{{d^2\Phi_u(t)}\over {dt^2}},$$
\end{equation}
and the inverse participation numbers $t_n$, $n\ge 2$,

\begin{equation}\label{IIIA2}
$$t_n\equiv\int^{\infty}_0t^nf(t)dt={{n(n-1)}\over V}\int_0^{\infty}
t^{n-2}\Phi_u(t)dt,$$
\end{equation}
can be related to the generating functional $\Phi_u(t)$ of the reduced
$\sigma$-model,

\begin{equation}\label{IIIA3}
$$\Phi_u(t)=\int_{\tilde {Q}({\bf r}_o)=\Lambda}{\rm D}\tilde {Q}(
{\bf r})e^{-F_u[t,\tilde {Q}]}.$$
\end{equation}
The free energy $F_u$ in Eq. (\ref{IIIA3}) has the form

\begin{equation}\label{IIIA4}
$$F_u[t,\tilde {Q}]=\int d{\bf r}{\rm S}{\rm t}{\rm r}\left[{{\pi\nu
D}\over 8}(\nabla\tilde Q)^2-{t\over 4}\Lambda\Pi\tilde Q({\bf r})\right
],$$
\end{equation}
and we remind that $\tilde {Q}({\bf r}_o)=\Lambda$ at the origin.  The
projection operator
$\Pi$ in Eq.  (\ref{IIIA4}) is defined as

\begin{equation}\label{IIIA41}
$$\Pi =\left(\matrix{\pi_b&\pi_b\cr
\pi_b&\pi_b\cr}
\right),\;\pi_b=\left(\matrix{0&0\cr
0&\tau_0\cr}
\right),$$
\end{equation}
and selects from the $Q$-matrix only its non-compact sector.

The generating functional $\Phi_u(t)$ has several funny features.  First, at
$t=0$, it has a completely invariant form, and, therefore, is equal to
unity, what corresponds to the normalization of the wave functions,

\begin{equation}\label{IIIA5}
$$Vt_1=\Phi_u(0)\equiv 1.$$
\end{equation}

On the other hand, for any finite $t$, the reduced $\sigma$-model is a model
with a broken symmetry, so that the free energy in Eq.  (\ref{IIIA4})
can be minimized by an inhomogeneous solution $\tilde {Q}({\bf r})$.  Indeed,
${
t\over 4}\Lambda\Pi$ in
the second term in Eq.  (\ref{IIIA4}) looks as a field tending to align
the matrix $\tilde {Q}$ along the non-compact 'direction' of the $
Q$-space (related
to the parameter $\theta_1$), whereas the boundary condition at ${\bf r}_
o$ together
with the gradient term is a rigidity attempting to prevent that.
The competition between these two tendencies results in an optimal
configuration of $\tilde {Q}$.  To find such an optimal (saddle-point)
solution,
we use, again, the invariance of the $Q$-space with respect to
rotations $\tilde {V}$ and represent $Q$ as

\begin{equation}\label{IIIA6}
$$\tilde {Q}({\bf r})=V_t({\bf r})\Lambda{{1+iP}\over {1-iP}}\bar {
V}_t({\bf r}),\;P=\left(\matrix{0&B\cr
\bar {B}&0\cr}
\right),$$
\end{equation}
where a weak perturbation $ $ $P$ stands for fluctuations around
the saddle-point, and the matrices $B,\bar {B}$ can be
decomposed into blocks as follows,

\begin{equation}\label{IIIA7}
$$B=\left(\matrix{s_{1,1}\tau_0+is_{1,2}\tau_3&\hat{\sigma}_1\cr
\hat{\sigma}_2^{+}&s_{2,1}\tau_0+is_{2,2}\tau_3\cr}
\right),$$
\end{equation}
$$\hat{\sigma}_{\alpha}=\left(\matrix{\sigma_{\alpha}&0\cr
0&-\sigma_{\alpha}^{*}\cr}
\right).$$

The form of the saddle-point $\tilde {Q}=V_t\Lambda\bar {V}_t$, follows from
the requirement
of the absence of linear terms in the expansion $F_u$ into the series

\begin{equation}\label{IIIA8}
$$F_u[t,\tilde {Q}]=F_t+F^{(2)}+F^{(3)}+F^{(4)}+...$$
\end{equation}
in the perturbation $P$.  This selects

\begin{equation}\label{IIIA9}
$$V_t=\exp\left(\matrix{0&{1\over 2}\theta_te^{i\chi\tau_3}\cr
{1\over 2}\theta_te^{-i\chi\tau_3}&0\cr}
\right),$$
\end{equation}
where the parameter $\theta_t({\bf r})$ satisfies the optimum equation

\begin{equation}\label{IIIA10}
$$\Delta\theta_t({\bf r})=-{t\over {\pi\nu D}}e^{-\theta_t};\;\chi
=\pi$$
\end{equation}
with the boundary conditions $\theta_t({\bf r}_o)=0$ in the origin and $
({\bf n}\nabla )\theta_t=0$ at
the surface of a sample.  In Eq. (\ref{IIIA10}) $\Delta$ stands for the
Laplacian in the real space.  This equation is partly similar to the
saddle-point equation derived by Muzykantskii and Khmelnitskii
\cite{Dima} when studying the problem of long-living current
relaxation in open conductors, but it has a different non-linearity and
- what is more important - different boundary conditions.

\subsection{Orthogonal ensemble}

The parametrization of $Q$-matrices in the orthogonal case is more
complicated due to a larger number of independent parameters in it.
In particular, the non-compact sector of the degeneracy space is
parametrized by two 'imaginary angles' - variables $\theta_{1,2}$:
$$\hat{\theta }=\left(\matrix{\theta\tau_0&0\cr
0&i(\theta_1\tau_0+\theta_2\tau_1)\cr}
\right),\quad\matrix{\,0<\theta <\pi ,\cr
0<\theta_{1,2}<\infty .\cr}
$$
Unitary matrices $u_2$ and $v_2$ in Eq.  (\ref{II8}) have a more
complicated form, too,
$$u_2=\left(\matrix{M&0\cr
0&e^{i\phi\tau_3}\cr}
\right),\;v_2=\left(\matrix{\tau_0&0\cr
0&e^{i\chi\tau_3}\cr}
\right),\;M={{1-i\vec {m}\vec{\tau}}\over {1+i\vec {m}\vec{\tau}}}
,$$
where $0\le\phi ,\chi <2\pi$, $m_{1,2,3}$ are real, and the number of
anti-commuting variables in $u_1$ and $v_1$ is twice as large as in the
unitary case.

The integration over $Q_o$ can be performed in a complete analogy with
the unitary case, but with several distinguishing features.  First of
all, in the limit of $\gamma\to 0$ the main contribution comes from the region
of the $(\theta_1,\theta_2)$-plane where $\cosh \theta_1\cosh \theta_
2\sim 1/\gamma$.  Since the product
$\cosh \theta_1\cosh \theta_2$ can be large at large $\theta_1$ as well as
large $
\theta_2$, we end
up with the new form of projection operators,
$$\Pi_o=\left(\matrix{\pi_b(o)&\pi_b(o)\cr
\pi_b(o)&\pi_b(o)\cr}
\right),\;\pi_b(o)=\left(\matrix{0&0\cr
0&\tau_0+\tau_1\cr}
\right),$$
in the free energy

\begin{equation}\label{IIIB1}
$$F_o[t^{\prime},\tilde {Q}]=\int d{\bf r}{\rm S}{\rm t}{\rm r}\left
[{{\pi\nu D}\over 8}(\nabla\tilde Q)^2-{{t^{\prime}}\over 8}\Lambda
\Pi_o\tilde Q({\bf r})\right].$$
\end{equation}
This $\Pi_o$ determines the direction of an effective 'force'
along the symmetrically chosen non-compact 'direction'
$(\theta_1+\theta_2)$.

Next, in the orthogonal ensemble, one has to keep the external phase
factor $e^{i\xi}$ in Eq.  (\ref{II10}) until the end of the integration over $
Q_o$,
which results in the integro-differential relation

\begin{equation}\label{IIIB2}
$$f_o(t)={4\over {V\pi\sqrt t}}{{d^2}\over {dt^2}}\left\{\int^{\infty}_
0dz\Phi_o(t+z^2)\right\}$$
\end{equation}
between the distribution function $f_o$ and generating functional

\begin{equation}\label{IIIB3}
$$\Phi_o(t^{\prime})=\int_{\tilde {Q}({\bf r}_o)=\Lambda}{\rm D}\tilde {
Q}\exp (-F_o[t^{\prime},\tilde {Q}]).$$
\end{equation}
The generating functional $\Phi_o(t^{\prime})$ gives directly the inverse
participation numbers $t_n,$ $n\ge 2$,

\begin{equation}\label{IIIB4}
$$t_n={2\over {\sqrt {\pi}V}}{{\Gamma (n+1/2)}\over {\Gamma (n-1)}}
\int_0^{\infty}(t^{\prime})^{n-2}\Phi_o(t^{\prime})dt^{\prime},$$
\end{equation}
and, again, has the property $\Phi_o(0)=Vt_1\equiv 1$.

To study the fluctuations near the saddle-point, we represent $\tilde {
Q}$ in the
form of Eq.  (\ref{IIIA6}), where

$$B={{B_{+}+B_{-}}\over {\sqrt 2}},$$
$$B_{\pm}=\left(\matrix{(s_{11}^{\pm}\tau_0+is_{12}^{\pm}\tau_3)\tau_{
\pm}&\hat{\sigma}_1^{(\pm )}\cr
(\hat{\sigma}_2^{(\pm )})^{+}&(s_{21}^{\pm}\tau_0+is_{22}^{\pm}\tau_
3)\tau_{\pm}\cr}
\right),$$
and
$$\;\;\hat{\sigma}_{\alpha}^{(\pm )}=\left(\matrix{\sigma_{\alpha}^{
\pm}&\pm\sigma_{\alpha}^{\pm}\cr
\mp (\sigma_{\alpha}^{\pm})^{*}&-(\sigma_{\alpha}^{\pm})^{*}\cr}
\right),\;\tau_{\pm}=\tau_0\pm\tau_1.$$
In this decomposition, $s^{\pm}_{\alpha\beta}$ are real numbers, $
\sigma_{\alpha}^{\pm}$ and $(\sigma_{\alpha}^{\pm})^{*}$ -
anti-commuting variables.  Indices '$\pm$' are introduced for the later
convenience.  Everywhere below, we keep superscript '-' but omit '$
+$'.

Similarly to the unitary case, the free energy $F_o$ which governs the
generating functional $\Phi_o$ has the minimum at $\tilde {Q}=V_t\Lambda
\bar {V}_t$,
$$V_t=\exp\left(\matrix{0&e^{i\chi\tau_3}\sum_{\pm}{{\theta_t^{\pm}
\tau_{\pm}}\over 4}e^{-i\phi\tau_3}\cr
e^{i\phi\tau_3}\sum_{\pm}{{\theta^{\pm}_t\tau_{\pm}}\over 4}e^{-i\chi
\tau_3}&0\cr}
\right)$$
where the variables $\theta_t\equiv\theta_1+\theta_2$, $\theta_t^{
-}\equiv\theta_1-\theta_2$ and $\chi ,\phi$ satisfy the
following equations:

\begin{equation}\label{IIIB5}
$$\Delta\theta_t({\bf r})=-{{t^{\prime}}\over {\pi\nu D}}e^{-\theta_
t},\;\Delta\theta_t^{-}({\bf r})=0,\;e^{i(\chi\pm\phi )}=-1.$$
\end{equation}
The conditions at the origin and boundary are the same, as in
the unitary case.  The latter gives $\theta_t^{-}({\bf r})=0$, and the
non-trivial
saddle-point is related only to the symmetric variable $\theta_t$.

\subsection{Symplectic ensemble}

An analogous investigation of the statistics of spin-polarized electron
waves in the case of a strong spin-flip scattering needs an extension
of dimensions of $Q$-matrices and the following analysis of the
degeneracy space related to their gapless Goldstone modes.  The gaps
in the spectrum of $Q$'s appear due to a large spin-relaxation rate, $
\tau_s^{-1}$,
which can be caused both by the spin-orbit coupling built into the
material properties or by the spin-flip scattering on a classical
randomly oriented static magnetic field.  In the former case, the
time-reversal symmetry is conserved, while in the latter this
invariance is violated by the source of a scattering.  Since triplet
components of $Q$ correspond to gapfull modes \cite{WeakLoc,Efetov},
only singlet modes have to be taken into account, so that the number
of independent variables of $Q$ is the same as in the spinless case.
One has to remember only that all the matrix elements of $Q$ are
multiplied by the unit $2\times 2$ spin-matrix $\tilde{\tau}_0$.

In this subsection, we work with the distribution $f(\downarrow ,t
)$ of a local
spin density $t\equiv |\psi_{\downarrow}|^2$ of the eigenstates defined by Eq.
(\ref{IA5})
and above it.  An incorporation of spins into Eq.  (\ref{II10}) can be
done by substituting
$$\pi_{1,2}\to\pi_{1,2}\otimes\tilde{\tau}_0,\;\Upsilon\to\Upsilon
\otimes\tilde{\tau}_{\downarrow},$$
where '$\otimes$' stands for the direct product of matrices, and $
\tilde{\tau}_i$ are the
Pauli spin operators:  $\tilde{\tau}_0=\left(\matrix{1&0\cr
0&1\cr}
\right)$, $\tilde{\tau}_{\downarrow}=\left(\matrix{0&0\cr
0&1\cr}
\right)$.
The degeneracy space of $Q_o$ is non-compact along a single
direction, and the integration over $Q_o$ gives us

\begin{equation}\label{IIIC1}
$$f(\downarrow ,t)={1\over {2V}}{{d^2\Phi_s(t)}\over {dt^2}},\quad
\Phi_s(t)=\int {\rm D}\tilde {Q}({\bf r})e^{-F_s[t,\tilde {Q}]},$$
\end{equation}
where

\begin{equation}\label{IIIC2}
$$F_s[t,\tilde {Q}]=\int d{\bf r}{\rm S}{\rm t}{\rm r}\left[{{\pi\nu
D}\over 4}(\nabla\tilde Q)^2-{t\over 2}\Lambda\Pi\tilde Q({\bf r})\right
],$$
\end{equation}
and $\Pi$ is exactly the same as in Eq.  (\ref{IIIA41}).

 From the point of view of the rest of calculations, the case with a
broken time-reversal invariance is equivalent to the spinless unitary
symmetry class \cite{Efetov}.  That's why we mark the quantities
related to this symmetry with a label '$su$' and generate the
distribution function $f_{su}(\downarrow ,t)$ form the distribution function $
f_u(t)$ in
Eqs.  (\ref{IIIA1}-\ref{IIIA4}) as
$$f_{su}(\downarrow ,t)=2f_u(2t)\;{\rm a}{\rm t}\;D\to 2D.$$

As concerns the symplectic ensemble with the time-reversal
symmetry, it demands an extra calculation, since in the presence of
the spin-orbit interactions it cannot be reduced to the spinless
orthogonal one.  The parametrization of the $Q$-space in this case is
given by Eqs.  (\ref{II8}) with
$$\hat{\theta }=\left(\matrix{\theta\tau_0+\theta^{\prime}\tau_1&0\cr
0&i\theta_1\tau_0\cr}
\right),\quad\matrix{\,0<\theta ,\theta^{\prime}<\pi ,\cr
0<\theta_1<\infty ,\cr}
$$
and
$$u_2=\left(\matrix{e^{i\phi\tau_3}&0\cr
0&M\cr}
\right),\;v_2=\left(\matrix{e^{i\chi\tau_3}&0\cr
0&\tau_0\cr}
\right),\:M={{1-i\vec {m}\vec{\tau}}\over {1+i\vec {m}\vec{\tau}}}$$
where $0\le\phi ,\chi <2\pi$, and $m_{1,2,3}$ are real variables.  After this,
the
saddle-point configuration of $\tilde {Q}$ for the symplectic-orthogonal case
can
be found as $\tilde {Q}=V_t\Lambda\bar {V}_t$,
$$V_t=\exp\left(\matrix{0&{{\theta_t}\over 2}M\cr
{{\theta_t}\over 2}M^{+}&0\cr}
\right),$$
where $\theta_t$ satisfies the optimum equation and the matrix $M$ is chosen
in such a way that $M=-\tau_0$:

\begin{equation}\label{IIIC3}
$$\Delta\theta_t({\bf r})=-{t\over {\pi\nu D}}e^{-\theta_t},\;|m|\to
\infty ,$$
\end{equation}
with the boundary conditions $\theta_t({\bf r}_o)=0$ and $({\bf n}
\nabla )\theta_t=0$ at the surface.

\subsection{Optimal free energy and fluctuations near the saddle-point}

After comparing the saddle-point equations in Eqs.
(\ref{IIIA10},\ref{IIIB5},\ref{IIIC3}), one finds that they are similar in
different symmetry classes.  The difference between the unitary,
orthogonal and symplectic ensembles leads only to different values of
a coefficient $\beta$,

\begin{equation}\label{IIID0}
$$\beta_o={1\over 2},\;\beta_u=1,\;\beta_{su}=\beta_s=2,$$
\end{equation}
in the expression for the optimal free energy

\begin{equation}\label{IIID1}
$$F_t=\beta\int d{\bf r}\left\{{{\pi\nu D}\over 2}(\nabla\theta_t)^
2+te^{-\theta_t}\right\},$$
\end{equation}
and in the higher order terms of the expansion of $F[t,\tilde {Q}]$ in the
environs of the saddle-point.  The generating functional $\Phi (t)$ from Eqs.
 (\ref{IIIA3},\ref{IIIB3},\ref{IIIC1}) can be represented in the form
$$\Phi (t)=J(t)\exp (-F_t).$$
In the conducting regime, the value of the optimal free energy
determines the leading term in the exponential of the generating
functional, whereas the effect of fluctuations around the saddle-point
is included into the function $J(t)$,

\begin{equation}\label{IIID2}
$$J(t)=\int {\rm D}P\exp \{-F^{(2)}-F^{(3)}-F^{(4)}...\}.$$
\end{equation}

Due to the normalization condition in Eq.  (\ref{IIIA5}), the relation
$J(0)=1$ holds exactly, and the contribution from the fluctuations $
P$
can be calculated by expanding the exponential in the integrand in Eq.
(\ref{IIID2}) into the series in the higher-order terms $F^{(3,4,.
..)}$ and
performing Gaussian integrations over with the weight $\exp \{-F^{
(2)}\}$
determined by the second-order correction to the free energy.

The applicability of such a perturbation theory is justified by the
fact that the higher orders are, at least, by the factor of $(2\pi
\nu D)^{-1}\ll 1$
smaller, as compared to what is given by

\begin{equation}\label{IIID3}
$$J(t)\approx\int {\rm D}P\exp \{-F^{(2)}[t,P]\}.$$
\end{equation}
The latter is nothing but the super-determinant of the Hamiltonian
related to the fluctuations around the saddle-point.  The value of $
J(t)$
differs from unity merely because the symmetry between fermionic
and bosonic degrees of freedom is broken by the optimal solution.
Since not all the projections of the infinitesimal $P$ to the generators
of the Lie algebra of the graded symmetry group are equivalently
affected by the symmetry breaking, it is convenient to separate in
$F^{(2)}$ the terms which feel the existence of the optimal solution from
those which do not.  Depending on the physical symmetry class, this
involves different sets of variables.  Nevertheless, after an
appropriate diagonalization, quadratic form $F^{(2)}$ can be represented
uniquely for all symmetry classes:

\begin{equation}\label{IIID4}
$$F^{(2)}=F^{(2)}_t+\varsigma F^{(2)}_0,\;\left\{\matrix{\varsigma_{
u,su}=0\cr
\varsigma_{o,s}=1\cr}
\right..$$
\end{equation}
The term $F_0^{(2)}$ in Eq. (\ref{IIID4}) is composed of fields which are
not affected by the symmetry breaking,
$$F^{(2)}_0=2\pi\nu D\sum_{\alpha =1,2}\int d{\bf r}\left\{\vec\partial
\sigma_{\alpha}^{-}\vec\partial (\sigma_{\alpha}^{-})^{*}+\vec\partial
s_{\alpha}^{-}\vec\partial (s_{\alpha}^{-})^{*}\right\}.\;$$
This term does not contribute to the function $J$ in the Gaussian
approximation, due to the symmetry between boson and fermion
degrees of freedom incorporated in it.  On the contrary, the first
term in Eq.(\ref{IIID4}) is the sum over those four pairs of dynamical
variables which feel the violation of the boson-fermion symmetry
$$F^{(2)}_t=2\pi\nu D\int d{\bf r}\left\{\sum_{\alpha =1,2}[\partial
\sigma_{\alpha}\partial\sigma_{\alpha}^{*}+U_{\sigma}\sigma_{\alpha}
\sigma_{\alpha}^{*}]\right.$$
$$\left.+\sum_{\beta ,\alpha =1,2}[(\partial s_{\alpha ,\beta})^2+
U_s^{\alpha\beta}s_{\alpha ,\beta}^2]\right\}.$$
Due to that, the pre-exponential $J$ can be represented as

\begin{equation}\label{IIID5}
$$J=\exp\left\{{1\over 2}\sum_n\ln\left((\chi_{\sigma}(n))^4/\prod_{
\alpha ,\beta}\chi_s^{\alpha\beta}(n)\right)\right\},$$
\end{equation}
where the sum is extended over all the eigenvalues of the spectral
problem

\begin{equation}\label{IIID6}
$$[-\Delta +U-\chi ]\phi =0,\quad\phi ({\bf r}_o)=0,\;{\bf n}\nabla
\phi (S)=0.$$
\end{equation}

As we already mentioned above, in the quadratic approximation, any
difference of $J$ from unity is due to the broken symmetry between
fermionic and bosonic degrees of freedom in $P$.  The broken symmetry
in the Hamiltonian $F_t^{(2)}$ is the consequence of the difference between
the effective potentials

\begin{equation}\label{IIID7}
$$\matrix{U_{\sigma}={1\over 4}(\nabla\theta_t)^2+{t\over {2\pi\nu
D}}e^{-\theta_t},\cr
U_s^{\alpha\beta}={{k_{\alpha ,\beta}}\over 4}(\nabla\theta_t)^2+{{
tq_{\alpha ,\beta}}\over {2\pi\nu D}}e^{-\theta_t}\cr}
\;$$
\end{equation}
in the Hamiltonian $F_t^{(2)}$.  In Eq. (\ref{IIID7}), $k_{\alpha
,1}=q_{\alpha ,1}=0$ and
$k_{1,2}=4$, $k_{2,2}=0$, $q_{\alpha ,2}=2$.  The spectra $\{\chi_{
\sigma}(n)\}$ and $\{\chi_s^{\alpha\beta}(n)\}$ of
modes remain gapfull, since all $U>0$.  Moreover, due to the sum rule

\begin{equation}\label{IIID8}
$$\sum_{\alpha\beta}U_s^{\alpha\beta}=4\sum U_{\sigma},$$
\end{equation}
their main contribution to $J$ comes from low-lying eigenvalues of Eq.
(\ref{IIID6}).  Since the set of $U$'s in Eq.  (\ref{IIID7}) depends
on the optimal solution alone, the calculation of the correction to the
exponent in this order in $(2\pi\nu D)^{-1}$ can be performed simultaneously
for all symmetry classes and is small.

On the other hand, the effect of fluctuations can become important
once we want to extend the consideration of samples with the size
larger than the localization length.  Such a step which is left beyond
of the scope of this paper would need some kind of a renormalization
of the saddle-point.  We would only like to stress that this can be a
way to avoid the previously found difficulty
\cite{KLY,Chakr,Wegner2} to obtain a stable fixed-point in the
renormalization group treatment of the localization problem.

The existence of the saddle-point and relatively small contribution
from fluctuations in the metallic regime makes it easy to find the
form of the cross-correlation function $R(t,r)$ from Eq.  (\ref{IA5}),
too.  If we study the envelope of the wave function at large enough
distances $r\gg l$ from the position of a high amplitude splash, the
reasoning used above can be repeated for $R(t,r)$ with minor
modifications, and we arrive at

\begin{equation}\label{IIID9}
$$R(t,r)/f(t)\propto te^{-\theta_t(r)},$$
\end{equation}
so that one can say that the envelope of $|\psi (r)|^2$ follows the form of
the saddle-point configuration of the reduced $\sigma$-model.

Up to now, we performed our analysis without referring to any
specific dimensionality of the system.  On the other hand, from Eqs.
(\ref{IIIA10},\ref{IIIB5},\ref{IIIC3},\ref{IIID6}), one can see that the
saddle-point solution, and, therefore, the optimal free energy $F_
t$
crucially depend on the dimensionality.  In the next Sections, we
discuss the statistics of local amplitudes of wave functions in 1D, 3D
and 2D conductors separately.

\section{Nearly localized states in a short disordered wire}

It is well known that the localization effects are the strongest in
one- and quasi-one-dimensional (1D) conductors
\cite{Thouless,WeakLoc}.  Even when disorder is weak, the quantum
diffusion of a particle is blocked at the length scale $L_c=\beta
2\pi\nu D$,
where $D$ is the classical diffusion coefficient determined by the
elastic impurity scattering \cite{Efetov} and $\beta$ is specified by Eq.
(\ref{IIID0}).  Since in the quasi-1D wire the effective density of
states $\nu$ is that integrated over cross-sectional width $w$ or an area,
$w^2$, the localization length $L_c\sim l(w/\lambda_F)^{d-1}$ can be much
longer than
the mean free path $l$.  This allows us to consider the short wires
$L<L_c$ with an already developed diffusive regime, and below we
analyze the distribution of amplitudes and the shape of untypical
states which are the precursors of localization at larger distances.
Since there is a lot of known about the 1D case
\cite{Thouless,EfetovLarkin,MF}, its example can be a good point to
compare the results of the saddle-point approach with calculations
based on the use of the exact transfer-matrix method \cite{Feynman}.
In fact, in the 1D case, the saddle-point calculation is nothing but the
'semiclassical' solution of the effective Schroedinger equation on the
$Q$-space \cite{Efetov} which appears in the transfer-matrix method.
Such a 'semiclassics' not only gives the results which are very close
to the exact solutions \cite{MF}, but also enables us to make a
statement about the form of typical pre-localized states in the
metallic regime.

In the following paragraphs, we apply the scheme of calculus
described in the previous section.  First of all, we have to solve the
saddle-point equation,

\begin{equation}\label{IV1}
$$\partial^2_x\theta_t(x)=-{t\over {\pi\nu D}}\exp \{-\theta_t\},$$
\end{equation}
and use its solution $\theta_t(x)$ when calculating the optimal free energy
$F_t$.  Due to the condition $\theta_t(x_o)=0$ at the observation point $
x_o$, the
latter  splits the wire with the length $L$ into two (not necessarily
equal)  intervals $0<x<L_{i=L,R}$.  The form of $\theta_t(x)$ can be found
separately in each of them.  In dimension one, the differential
equation in Eq.  (\ref{IV1}) can be solved exactly \cite{Kamke}, and
we represent its general solution in the form

\begin{equation}\label{IV2}
$$e^{-\theta_t}=\left[{{A_i}\over {\cos\left\{A_i\sqrt {T_i}\left(
1-{x\over {L_i}}\right)\right\}}}\right]^2,\quad x>0;\:i=L,R.$$
\end{equation}
Although one can notice that the general form of this solution
formally contains a singularity at $x_{\infty}=-L_i[\pi /(2A_i\sqrt {
T_i})-1]$, the latter
is illusory since it takes place in the non-physical region $x<0$ and
plays no role unless it comes up to the formulation of the
limitations to our theory.  The requirement $x_{\infty}\gg l$ which emerges
from the existence of the singular point is related to the conditions
on maximal values of gradients permitted by the use of only the
lowest-order gradient expansion terms in the free energy functional
in Eq.  (\ref{II10}).  We shall discuss the consequences of this
condition at the end of the Section, assuming for a while that it is
satisfied.  If so, the consistency equations on the parameters $A_
i$
come from the boundary conditions $\partial_x\theta_t(L_i)=0$ at the edges and
have
the form

\begin{equation}\label{IV3}
$$A_i=\cos\left(A_i\sqrt {T_i}\right),$$
\end{equation}
where $T_i$ are defined as

\begin{equation}\label{IV4}
$$T_i={{tL_i^2}\over {2\pi\nu D}}.$$
\end{equation}
The optimal value of the free energy can be found, in its turn, as

\begin{equation}\label{IV5}
$$F_t=\beta\sum_{i=L,R}{{L_it}\over {\sqrt {T_i}}}\left[2\sqrt {1-
A_i^2}-A_i^2\sqrt {T_i}\right].$$
\end{equation}

In general, the exact form of $F_t$ in Eq.  (\ref{IV5}) based on the
closed set of equations in Eqs.  (\ref{IV2}-\ref{IV4}) can be studied
numericaly at any values of the parameters included, but a somewhat
simpler analytical expression can be written in the asymptotical
regions.  First of all, we examine the limit of small amplitudes,
$T_i<1$, where the exact distribution has to match with the random
matrix theory results.  At $T_i<1$, the results of Eqs.
(\ref{IV3}-\ref{IV5}) can be expanded into the series in $T_i$, which
gives $A_i\approx 1-{1\over 2}T_i+{{13}\over {24}}T_i^2+...$  and
$$F_t\approx Vt\left(1-\sum_{i=L,R}{{T_iL_i}\over {3L}}+...\right)
,\quad f^{(1)}(t)\approx Ve^{-\beta F_t}.$$
We see that in this limit the Porter-Thomas formulae, Eq.
(\ref{IB1}-\ref{IB3}) give a good approximation for the body of the
distribution function $f(t)$ which describes those amplitudes $t$ which
are $t<L^{-1}\sqrt {L_c/L}$.  Otherwise, the second term of this expansion, $
VtT_i$,
becomes larger than unity and strongly affects the probability to
detect a too high splash of the wave function.

When $T_i\gg 1$, the optimal configuration $e^{-\theta_t}$ develops at the
length
scale of $\lambda =\sqrt {2\pi\nu D/t},$ where it can be approximated as

\begin{equation}\label{IV6}
$$e^{-\theta_t(x)}\sim (\lambda /x)^2,$$
\end{equation}
and gets less sensitive to the presence of boundaries.  Indeed, in the
limit of $T_i\gg 1$, one has $A_i\approx{{\pi}\over 2}T_i^{-1/2}(1
-T_i^{-1/2}+...)$, and the exact
expression for the optimal free energy can be expanded into the
series in $T_i^{-1/2}$,

\begin{equation}\label{IV7}
$$F_t=4\sqrt {\beta L_ct}\left\{1-\delta_L-\delta_R\right\},$$
\end{equation}
where
$$\delta_i\sim{{\pi^2}\over 8}[T_i^{-1/2}-{1\over 2}T_i^{-1}+...],
\quad i=L,R.$$
The leading term in Eq.  (\ref{IV7}) does not depend on the system
length and (in the orthogonal ensemble) coincides with the
asymptotical behavior of the distribution function of local amplitudes
in infinite wires.  The latter has been found by Mirlin and Fyodorov
\cite{MF} on the basis of the analysis of the transfer-matrix
equations derived by Efetov and Larkin \cite{EfetovLarkin}.  Although
we did our calculations for the short-length samples, $L<L_c=\beta
2\pi\nu D$,
the results for the high amplitude splashes surprisingly agree with
those for the infinite geometry, even up to the leading term of
pre-exponential factor $J$.  The latter is derived in Appendix A by
taking into account fluctuations near the saddle-point.  Roughly
speaking, the 'semiclassical' solution of the transfer-matrix equation
gives an almost exact result.  The full form of the tails of $f(t)$ at
$t>g/L$, where $g=L_c/L$, can be represented as

\begin{equation}\label{IV8}
$$f^{(1)}(t)\approx C{{\sqrt {L_LL_R}}\over L}\sqrt {{{L_c}\over t}}\exp\left
[-4\sqrt {\beta L_ct}\left\{1-\delta_L-\delta_R\right\}\right]$$
\end{equation}
and is applicable up to the amplitudes $t\sim L_c/l^2\sim w/(l\lambda_
F)$.  The
latter condition emerges from the requirement of a smoothness of
the saddle-point solution, so that its characteristic length scale $
\lambda$
should be longer than the mean free path, $\lambda >l$.  Otherwise, the
singularity of the equation in Eq. (\ref{IV2}) comes too close to the
physical part of the space ($x>0$) which would create too large
gradients forbidden within the framework of the used approximations.

The distribution function given by Eq.  (\ref{IV8}) evidences that the
states which are responsible for the rare event we discussed in the
previous paragraph are (at least, partly) localized.  Nevertheless,
even for the largest amplitudes $t>L_c/L^2$, the effect of edges is still
present, which means that this in not an exponential localization.  On
the basis of an analysis of the cross-correlation function $R(t,r)$ from
Eqs.  (\ref{IA5}) and (\ref{IIID9}), we can say that, within the range
of distances $x<L_c$, the {\it envelope of the density of pre-localized
states $\psi_t(x)$\/} resembles the form of the optimal solution and has the
power-law intermediate asymptotics

\begin{equation}\label{IV9}
$$|\psi_t(x)|^2\propto te^{-\theta_t}\sim L_c/x^2.$$
\end{equation}
In contrast to the 2D case which we discuss in Section VII, the
derived form of a typical wave function has the same exponent for
all pre-localized states, independently of the amplitude of their top
splashes.  Further, the tails of the envelope in Eq.  (\ref{IV9}) are
integrable, so that the inverse participation numbers which one can
find on the basis of Eqs.  (\ref{IA3}) and (\ref{IIIA2},\ref{IIIB4}) do not
indicate any fractal behavior.

\section{Eigenstates statistics in $d=3$}

The localization effects in weakly disordered 3D conductors are
known to be the least pronounced \cite{WeakLoc}, so that the eigenstates
statistics of 3D conductors has to be most similar to the universal
one.  Nevertheless, even in this case, not all states are described by
the Porter-Thomas distribution, and this Section is
devoted to the disorder-dependent corrections to formulae in Eqs.
(\ref{IB1}-\ref{IB3}) in dimension three.

To describe the statistics beyond the universality limit, we have to
solve optimum equations (\ref{IIIA10},\ref{IIIB5},\ref{IIIC3}).  For the
sake of simplicity, we consider a spherically shaped conducting
particle and place the observation point into its center.  This enables
us to seek for the solutions $\theta_t(r)$ in a symmetric form.  Nonetheless,
even that does not help us to find the exact form \cite{Kamke} of
a general solution of the non-linear equation

\begin{equation}\label{V1}
$$\Delta\theta_t(r)=[r^{-2}\partial_rr^2\partial_r]\theta_t=-{t\over {
\pi\nu D}}e^{-\theta_t},$$
\end{equation}
so that we have to develop the following approximate procedure.
The non-linear $\sigma$-model, Eq.  (\ref{II9}), was derived under the
conjecture of smoothly varying $Q$-fields.  This implies that the
distances shorter than the mean free path $l$ are excluded from our
consideration, and the condition $\theta_t(0)=0$ at the origin has to be
substituted by the condition $\theta_t(r_0)=0$ at the sphere of a radius
$r_0\sim l$.  After this, we scale the distances by $l$, so that $
\theta_t=\theta_t(r/l)$,
and solve the problem iteratively.  The iterative procedure appeals to
the fact that the Laplace equation which one can get by neglecting
the right hand side of Eq.  (\ref{V1}) has non-zero solutions and that
the parameter
$$\rho =\sqrt {\pi\nu D/(l^2t)}$$
which appears after rescaling the distances $r$ with the mean free
path is large.  The latter condition restricts our consideration to the
amplitudes $t<(l\lambda_F^2)^{-1}$ which are smaller than the density formed by
the forward-and-backward scattered trajectory between two
impurities.

As the first step, we expand the desired function $\theta_t(r)$ as
$$\theta_t\approx\theta_t^{(0)}+\theta_t^{(1)},\quad\theta_t^{(0)}
=A(1-l/r).$$
The first term in it satisfies the Laplace equation but does not
satisfy the necessary boundary condition at $r=L$.  The term $\theta_
t^{(1)}$ is
added in order to satisfy the requirement $\partial_r\theta_t=0$ at the
external
edge.  It must turn to zero at $r=l$ and can be found from the
linearized equation
$$[u^{-2}\partial_uu^2\partial_u]\theta_t^{(1)}=-\rho^{-2}\exp (-\theta_
t^{(0)}),\quad\theta_t^{(1)}(1)=0,$$
where $u=r/l$.  After this, the non-linearity of Eq.  (\ref{V1})
transforms into a self-consistent determination of the parameter $
A$
from the algebraic equation
$$A=\rho^{-2}\int_1^{L/l}u^2du\exp \{-A(1-u^{-1})\}\approx{1\over {
3\rho^2}}\left({L\over l}\right)^3e^{-A},$$
which arises from the requirement $\partial_u\theta_t^{(1)}(L/l)=0$.  One could
continue the iterative scheme even further and add corrections which
improve the function $\theta_t^{(1)}(x)$ itself, and so on, but this is not
necessary for evaluating the leading terms of the optimal free energy
$F_t$.  So, we stop the iteration after the first step and find that

\begin{equation}\label{V2}
$$e^{-\theta_t(r)}=\exp\left(-A(1-l/r)\right),\;r>l.$$
\end{equation}

The combination of the parameters which stands in the right-hand
side of Eq.  (\ref{V1}) and the self-consistency equation itself can be
rewritten in the form

\begin{equation}\label{V3}
$$\;Ae^A=T\equiv{{Vt\eta_3}\over {2\pi^2\nu D}}\sim Vt/(p_Fl)^2,$$
\end{equation}
where $\eta_3\sim (2l)^{-1}$ and the condition $p_Fl\gg 1$ corresponds to the
limit
of a weak disorder.   The optimal free energy related to the
saddle-point configuration can be calculated, too, and has the form

\begin{equation}\label{V4}
$$F_t\approx\beta 2\pi^2{{\nu D}\over {\eta_3}}A\left\{1+{A\over 2}\right
\}.$$
\end{equation}

When $T\ll 1$, the calculation both of the self-consistent $A$ and the
related value of $F_t$ can be performed as an expansion into a series in
the parameter $T$, i.e.,  we approximate $A\approx T-T^2+...$ and
$$F_t=\beta Vt\left[1-{{Vt\eta_3}\over {4\pi^2\nu D}}+...\right],\quad
\eta_3\sim (2l)^{-1}.$$

When $T\gg 1$, the leading terms arise from the estimation $A\sim\ln
T$.  In
each of these two regimes, the generating functional $\Phi (t)$ has the
form

\begin{equation}\label{V5}
$$\Phi^{(3)}\approx\left\{\matrix{\exp\left(-\beta Vt+\beta{{(Vt)^
2\eta_3}\over {4\pi^2\nu D}}+...\right),\;T<1,\cr
\exp\left\{-\beta\pi^2{{\nu D}\over {\eta_3}}\ln^2T+...\right\},\;
T>1,\cr}
\right.$$
\end{equation}
which can be used for evaluating both the distribution function
$f(t)$  and IPN's using Eqs. (\ref{IIIA1},\ref{IIIB2},\ref{IIIC1}) and
(\ref{IIIA2},\ref{IIIB4}).  At this point, we have to mention that the
coefficient $\eta_3\sim (2l)^{-1}$  cannot be determined better than by the
order
of magnitude.  We also remind that different symmetry classes are
distinguished by the parameter $\beta$: $\beta_u=1$, $\beta_o={1\over
2}$, $\beta_s=2$.

Eq.  (\ref{V5}) indicates that the noticeable deviations from the
universal Porter-Thomas distribution start rising at local densities
$tV\sim p_Fl$ (the second term in the exponent in Eq.  (\ref{V5}) becomes
larger than unity) and then develop into the logarithmically-normal
asymptotics at $tV>(p_Fl)^2$.  On one hand the states which generate
such an asymptotical tail are not typically metallic.  On the other
hand, both the form of the envelope,

\begin{equation}\label{V6}
$$|\psi_t(r)|^2\propto\exp \{-A(1-l/r)\},$$
\end{equation}
which we extract from the shape of the optimal solution in Eq.
(\ref{V2}) and the scaling of IPN's with the integer power of the
system volume for any $n$,

$$t_n\approx{{\min \{\varphi (n),[2\pi^2\nu D/\eta_3]^n\}}\over {V^
n}}\exp\left({{n^2\eta_3}\over {\beta 4\pi^2\nu D}}\right),$$
$$\varphi_u=n!,\;\varphi_o=(2n-1)!!,\;\varphi_s=n!/2^n,$$
indicate that these 3D states are not localized in a conventional
sense:  They always have a finite part of the density 'equally' smeared
all over the sample.  Of course, these extended density-tails decrease
when $t$ approaches the limiting value $t\sim (l\lambda_F^2)^{-1}$, but our
methods do
not allow us to make a statement about the structure of standing
waves at the scale of $r<l$.

The version of the $\sigma$-model we used above also restricts our
consideration to the metallic regime $p_Fl\gg 1$.  The development of a
theory at critical conditions, $p_Fl\le 1$, requires the use of more
sophisticated methods \cite{Efmedium}.  Nevertheless, the common
believe which arises from the most of known localization theories
\cite{WeakLoc} is that the dimension $d=2$ is critical, so that the
analysis of wave function statistics in 2D disordered conductors
would manifest the important features of the criticality.

\section{Multifractality of eigenstates in weakly disordered 2D
conductors}

To find an optimal configuration in the 2D case we limit the
length scale of the $Q$-field variations from below by the value of a
mean free path - similarly to what we discussed in the previous
Section.  This modifies the boundary conditions into $\theta_t(r_0
)=0$ at
$r_0\sim l$.  For the sake of simplicity, we consider the sample in the
form of a disk (with the radius $L$) and place the observation point $
{\bf r}_o$
right in its center.  Then, we seek for an axially symmetric
solution $\theta_t(r)$ of the equation
$$(r^{-1}\partial_rr\partial_r)\theta_t=-{t\over {\pi\nu D}}e^{-\theta_
t}.$$

\subsection{Exact solution}

This can be done both exactly \cite{Kamke} or using an iterative
procedure developed in Section VI.  The exact solution can be
represented in the form

\begin{equation}\label{VI1}
$$e^{-\theta_t}=\left[{{2(l/r)^{1-A}\left[\sqrt {({1\over {A\rho}}
)^2+1}+1\right]}\over {\left[\sqrt {({1\over {A\rho}})^2+1}+1\right
]^2-({1\over {A\rho}})^2({r\over l})^{2A}}}\right]^2,$$
\end{equation}
where $\rho =\sqrt {{{2\pi\nu D}\over {tl^2}}}$, and $A$ has to be found from
the boundary condition
at the sample edge $r=L$,

\begin{equation}\label{VI2}
$$\sqrt {A^2+\rho^{-2}}+A={{(L/l)^A}\over {\rho}}\sqrt {{{1+A}\over {
1-A}}}.$$
\end{equation}
After substituting the saddle-point solution $\theta_t(r)$ from Eqs.
(\ref{VI1}) to Eq.  (\ref{IIID1}), we also find the optimal free energy,
$$F_t=\beta 4\pi^2\nu D\left\{\ln\left({{(L/l)^{(1+A^2)}}\over {\rho^
2[1-A^2]}}\right)+2(1-\sqrt {A^2+\rho^{-2}})\right\}.$$
Together with Eq.  (\ref{VI2}), the latter expression can be used for
the numerical analysis of the exponential of the distribution function.

The numerical analysis shows that the consistency equation in Eq.
(\ref{VI2}) has positive roots only if $\rho >\ln (L/l)\gg 1$, which provides a
reasonable limitation to the wave functions amplitudes we can study
using this method:  We have to restrict the density $t$ of a splash by
the value $(\lambda_Fl)^{-1}$ related to the density of an electron state bound
to
the forward-and-backward scattered trajectory between two
impurities.  At the same time, in the limit of $\rho\gg 1$, the roots of Eq.
(\ref{VI2}) can be approximated by $A=1-\mu$, where $\mu <1$.  The same
conditions gives us a possibility to replace the exact solution in Eq.
(\ref{VI1}) by

\begin{equation}\label{VI3}
$$e^{-\theta_t}\approx (l/r)^{2\mu},$$
\end{equation}
what means that there is an easier way to get a satisfactory
approximate solution of the saddle-point equation in $d=2$ similar to
what we did in $d=3$.

\subsection{Solution using iterations}

The result of Eq.  (\ref{VI3}) can also be derived using the iterative
scheme discussed in the Section VI.  Being approved by the strong
inequality $\rho\gg 1$, we solve, first, the linear Laplace equation by
choosing its solution in the form which satisfies the boundary
conditions at the origin,
$$\theta_t^{(0)}=2\mu\ln (r/l),$$
where the parameter $\mu$ will be the subject of the next iteration.
That's, we seek for such $\theta_t=\theta_t^{(0)}+\theta_t^{(1)}(r
/l)$ which satisfies the
boundary condition at the external edge, and where
$$[u^{-1}\partial_uu\partial_u]\theta_t^{(1)}=-2\rho^{-2}u^{-2\mu}
,\quad\theta_t^{(1)}(1)=0,\;u=r/l.$$
This gives $\theta^{(1)}_t$  in the form
$$\theta^{(1)}_t(u)={{\rho^{-2}}\over {2(1-\mu )^2}}\left(1-u^{2(1
-\mu )}\right).$$
The requirement $\partial_r\theta_t(L)=0$ gives rise to the consistency
equation
and enables us to formulate the approximate procedure in the closed
form,
$$2\mu ={{(L/l)^{2-2\mu}}\over {(1-\mu )\rho^2}}.$$
The use of the iterative procedure is formally limited by a
requirement $\theta_t^{(1)}\ll 1$.

The parameter $\mu$ in the above equation can be found (with the
accuracy controlled by $1/\ln (L/l)\ll 1$) as

\begin{equation}\label{VI4}
$$\mu ={{z(T)}\over {2\ln (L/l)}},\;ze^z=T\equiv{{tV\ln (L/l)}\over {
2\pi^2\nu D}}$$
\end{equation}
and varies when the amplitude $t$ changes.  For example, the
cross-over of the optimal solution to the homogeneous $\tilde {Q}\equiv
\Lambda$ occurs
in the limit of $T\ll 1$ where one can find that
$$\mu ={1\over 2}T(1-T+...)/\ln{L\over l}$$
In the opposite limit of large amplitudes, $T\gg 1$,
$$\mu\approx{1\over 2}\ln T/\ln{L\over l}<1.$$

The approximate form of the optimal free energy can be found, in its
turn, as

\begin{equation}\label{VI5}
$$F_t\approx\beta 4\pi^2\nu D\{\mu +\mu^2\ln{L\over l}\}.$$
\end{equation}
When $T\ll 1$, this can be expanded in $T$ as
$$F_t\approx\beta Vt\{1-{T\over 2}+...\}.$$
When $T\gg 1$ (but still $t\ll (l\lambda_F)^{-1}$), the leading term in the
optimal $
F_t$
takes the form
$$F_t\approx\beta\pi^2\nu D{{\ln^2T}\over {\ln (L/l)}}.$$
Although the size of the system enters these formulae, the
logarithmically weak dependence of $F_t$ on $L$ makes it meaningful to
use the derived expressions for an arbitrary position of the
observation point inside the sample of an arbitrary convex shape.

The fluctuations around the saddle-point configuration and the
resulting pre-exponential factor $J(t)$ for the 2D case are discussed in
the Appendix B.  All over the conduction regime, their contribution to
the value of the generating functional is small, as compared to that
of the optimal solution itself.

\subsection{Distribution function and IPN's}

All this enables us to calculate the distribution function $f$.  For small
amplitudes $t<2\pi\nu D/[L^2\ln{L\over l}]$, one obtains

\begin{equation}\label{VI6}
$$f^{(2)}\approx V\exp\left(-\beta Vt\left[1-{T\over 2}+...\right]\right
)\times\left\{\matrix{\sqrt {{1\over {2\pi tV}}},\;o\cr
1,\;u\cr
2,\;s;su\cr}
\right.,$$
\end{equation}
where $T$ is defined in Eq.  (\ref{VI4}), and
$$\beta_o={1\over 2};\;\beta_u=1;\;\beta_{s,su}=2.$$
In the opposite limit, $t>2\pi\nu D/[L^2\ln{L\over l}]$,  the distribution
function takes
the form

\begin{equation}\label{VI7}
$$f^{(2)}\sim V\exp\left(-{{\beta\pi^2\nu D}\over {\ln (L/l)}}\ln^
2T\right).$$
\end{equation}
Eqs.  (\ref{VI6}) and (\ref{VI7}) generalize our earlier result
\cite{Multi} to various symmetry classes.  They show that for any
of the fundamental symmetries - orthogonal, symplectic and unitary -
disorder makes the appearance of high-amplitude splashes of wave
functions much more probable than one would expect from the
Porter-Thomas formula and, as concerns the most extra-ordinary
events, tends the tails to take the logarithmically-normal form.
When being written for the orthogonal ensemble, the
logarithmically-normal law in Eq.  (\ref{V5}) strikingly coincides with
the form of the asymptotics of the distribution of the local density
of states and conductance fluctuations in open systems found in Ref.
\cite{Alt}, although our theory was made for closed systems and is
based on a different scheme of calculations.  This agreement reveals
the deep relationship of these two results obviously caused by the
localization effects.  But the localization of wave functions which are
responsible for the asymptotic events is not the localization of a
particle in a confining potential.  The tails of these states do not
decay exponentially:  Even in the asymptotical regime $T\gg 1$, the size
$L$ of the system influences the distribution.  The splashes look as if
they were formed by focusing the waves by some rare configurations
of scatterers.  The structure of these states can be anticipated from
the way how their distribution feels the boundary or - directly -
from the cross-correlation functions $R(t,r)$ in Eqs.
(\ref{IA5},\ref{IIID9}).  Following the form of the optimal solution, the
envelope of the density of such a state has a power-law asymptotic
tail

\begin{equation}\label{VI8}
$$|\psi_t({\bf r})|^2\sim e^{-\theta_t(r)}\approx (l/r)^{2\mu}$$
\end{equation}
which approaches the limiting $r^{-2}$ dependence for the highest
amplitudes $t\sim (l\lambda_F)^{-1}$.

Moreover, the form of IPN's, $t_n$ derived on the basis of Eqs.
(\ref{IIIA2},\ref{IIIB4}) shows such a scaling with the size of a
system which implies them a multifractal structure.  To find the
moments $t_n$ accurately enough, we have to take into account that,
although the cross-over to the 0D case looks like the formal limit
$T(t)\to 0$, the Porter-Thomas statistics fails unless the condition
$tV\ll\sqrt {2\pi\nu D}$ is satisfied (see Eq.  (\ref{VI6})).  Hence, only
first few
ratios $t_n$, $2\le n\ll\sqrt {2\pi\nu D}$, can be estimated using a finite
polynomial
expansion of $f(t)$ into the series in $T$, and their first terms reproduce
corrections to the universal statistics found pertubatively in Ref.
\cite{FM}.  Alternatively, we derive the higher order IPN's from Eqs.
(\ref{IA3},\ref{IIIA2},\ref{IIIB3}) using the saddle-point method.  The
moments $t_n$ calculated in both ways are in a good agreement with
each other and, in the leading order, can be represented as

\begin{equation}\label{VI9}
$$t_n\approx{{\min \{\varphi (n),[2\pi\nu D/\ln{L\over l}]^n\}}\over {
l^{2\delta}V}}\left({1\over V}\right)^{(n-1)d^{*}/2},$$
\end{equation}
where

\begin{equation}\label{VI10}
$$d^{*}(n)\approx 2-{{\beta^{-1}n}\over {~4\pi^2\nu D}}.$$
\end{equation}

As one can see from Eqs.  (\ref{VI9},\ref{VI10}), we end up with such
a volume-dependence of the inverse participation numbers $t_n$ that
manifests the multifractal behavior of quantum states, Eq.
(\ref{IC1}).  The multifractality seems to be the generic property of
2D disordered systems.  The multifractal dimensions in Eq. (\ref{VI10})
are calculated in the leading order on the inverse conductivity,
so that all over the metallic regime, the dependence of $d^{*}$ on $
n$ and
disorder is accurate enough and qualitatively agrees with numerical
results \cite{Schreiber}.  Due to the limitation $t<(l\lambda_F)^{
-1}$, the above
equations work at $n\le 2\pi\nu D$, so that $n-\delta >0$, and the fractal
dimensions $d^{*}$ in Eq.  (\ref{VI10}) are positive.

\section{Discussions}

Summarizing the results of the paper, we studied the manifestation
of precursors of localization among the eigenstates of isolated
disordered conductors with the size smaller than the localization
length.  In order to detect them, we analyzed the statistics of local
amplitudes of wave functions, $t\equiv |\psi |^2,$ and, at the tails $
t\gg V^{-1}$ found
strong deviations from the universal Porter-Thomas distribution (see
Eqs.  (\ref{IB1}-\ref{IB3})) associated with the typically extended-type
behavior.  The universal statistics equally describes the quantum
states of various classically chaotic systems; it depends on their
fundamental symmetry but not on the physical dimensionality or the
level of disorder.  Such a description can be successfully applied to
the most of the states (extended ones) in the metallic regime and
gives the body of the distribution function of their local amplitudes.
The deviations from the universal laws start rising at the amplitudes
$t\sim\sqrt g/V$ and finally develop into a completely different asymptotics
at $t\sim g/V$.  In dimensions $d=2$ and 3, the form of the asymptotics
is described by the logarithmically-normal tails in Eq.
(\ref{IC2},\ref{V5},\ref{VI7}), whereas in $d=1$ it has a stronger
dependence, $f\propto\exp \{-4\sqrt {\beta tL_c}\}$.

The scheme of calculus (see Sections III and IV) and the similarity
between our results for isolated systems and the asymptotics of
distributions of the local density of states and fluctuations of other
quantities in 1D \cite{MF} and 2D \cite{Alt} conductors indicate that
the above-mentioned long tails are strongly influenced by the
localization.  To answer the question, how the localization develops,
we can refer to the fact that the deviations from the Porter-Thomas
distribution appear as a so small number of events, $\propto\exp (
-\sqrt g)$, that
their occurrence near the Fermi level in a specific sample is a
typically mesoscopic phenomenon.  We interpret this as that the
rare top-amplitude splashes are not locally implicit as portions to
any state but represent very non-trivial configurations of waves
which can be found more and more often if the disorder increases.
The analysis of the cross-correlations $R(t,r)$ also indicates that the
states which are responsible for locally the highest amplitudes
$|\psi ({\bf r}_o)|^2>g/V$ have individually specific envelopes of their
decaying
density far away from the observation point ${\bf r}_o$.  In 1D and 2D
systems, the tails of the envelopes obey the power-law dependence
$|\psi ({\bf r}_o+{\bf r})|^2\propto r^{-2\mu}$.  In $d=1,$ its exponent is $
\mu =1$, so that the density
of these tails is perfectly integrable at long distances, and one could
speak about them as about nearly localized ones.  In dimension $d=
2$
the exponent $\mu$ is limited by $\mu (t)\le 1$ (so that it is not the
localization
in the usual sense) and is individual for the states with different
amplitudes of the top-amplitude splash.  Such a behavior of
pre-localized states in $d=2$ coexists with the typically multifractal
behavior of the inverse participation numbers which has been
observed earlier in various numerical simulations at the critical
conditions of the localization-delocalization transition
\cite{Aoki,Schreiber,Kramer,Ohtsuki,Evangelou,Jan}.  Unfortunately, at
the present stage we cannot approach close to the 3D Anderson
transition and to check the multifractality globally.  Nevertheless,
even in $d=3$ we find the non-trivial logarithmically-normal
asymptotic behavior of the statistics, although the states which seem
to be responsible for that are not localized.

The combination of the facts presented above forces us to suggest
that the details of the structure and unusual statistics of rare
pre-localized states which we discussed in the present paper have
something to do with the statistics of extraordinary multiply
self-crossing diffusive paths which anomalously often return to the
same spatial coordinates ${\bf r}_o$.  This means that the almost confined
states develop because of rare shortened classical trajectories which
form a closed loop not only in the real space, but also in the full
phase space, since they finally come to the same 'sell' $dpdx=$ $\hbar$
where they started.  For instance, the most dense configuration could
be formed by a forward-and-backward scattered trajectory of a
particle bouncing few impurities.  An additional argument which
supports this scenario relates to the conditions limiting the validity
of our theory.  Basing on the use of the $\sigma$-model, we have
necessarily to cut the linear length scale of the supermatrix $Q$-field
variation from below by the value of the mean free path.
Nevertheless, the densities which can be described in our approach
are limited by the value $1/(l\lambda_F^{d-1}$) instead of a naively expected
inverse volume $l^{-d}$.  This is only possible if the states which we
study are locally anysotropic at the fine scale of distances of about $
l$
and typically have a snake-like structure with the cross-sectional
width $\sim\lambda^{d-1}_F$.

\section{Acknowledgements}

We are grateful to B.  Altshuler, B.  Kramer, I.  Lerner, B.  Simons
and M.  Schreiber for useful discussions and thank P.  Fulde for
continuous encouragements during all the time we worked on this
problem.  We especially acknowledge the discussions with V.
Kravtsov and his comments on the manuscript which helped to
improve it.  One of the authors (VF) acknowledges a partial support
from NATO Collaborative Research Grant No.  921333.

\section{Appendix A: Pre-exponential factor for quasi-1D case}

In this Appendix, we show some details of calculations of the
pre-exponential factor $J$ for quasi-1D case.  Due to the condition
$\theta_t=0$ at $x=x_o$, the observation point splits the interval $
[0,L]$ into
two pieces, and the spectrum $\{\chi (n)\}$ of fluctuations around the
saddle-point solution can be found in each interval separately.
Therefore, we represent the pre-exponential as a product $J=J_LJ_R$
of contributions from the left- and right-hand-side intervals with
lengths $L_{L,R}$, where each of $J_{L,R}$ is determined by the eigenvalues of
the Schroedinger equation in Eq.  (\ref{IIID6}) with the
symmetry-breaking potentials
$$U_i=\left[kT_i+\kappa\left(\left[{{\sqrt {T_i}\pi /2}\over {1+\sqrt {
T_i}}}\right]/\sin\left\{{{\sqrt {T_i}\pi /2}\over {1+\sqrt {T_i}}}{
x\over {L_i}}\right\}\right)^2\right]L_i^{-2},$$
where $i=L,R$.  When $T\ll 1$ (in the paragraph below, we omit indices
$L$ and $R$), these potentials can be
treated perturbatively.  Their first-order corrections cancel due to
the sum rule from Eq.  (\ref{IIID8}), so that $J\approx 1+T^2\approx
1$.  When
$T\gg 1$, the same cancellation eliminates contributions from the
high-excitation eigenvalues $\chi >(\pi /2L)^2T$, so that the important
contribution comes from the low-energy part of the spectrum,
$\chi <(\pi /2L)^2T$, where one can approximate
$$U\approx (\pi /2L)^2[k+k^{\prime}/\sin^2(\pi x/2L)],\quad k^{\prime}
=k+\kappa .$$
Using this approximation, the spectral problem of 1D Eq.  (\ref{IIID6})
can be solved exactly.  To find the exact solution, one has to change
variables from $x$ to $y=\cot (\pi x/2L)$ and then seek for solutions in the
form $\phi =P_n(y^2)/(1+y^2)^{\delta (n)}$, where $P_n(y^2)$ are polynomials.
This
results in the set of eigenvalues $\chi (n)$, $n\ge 0$,
$$\chi (n)=(\pi /2L)^2\left\{\left[2n+1/2+\sqrt {k^{\prime}+1/4}\right
]^2-k\right\}.$$
Being substituted into Eq.  (\ref{IIID5}), this gives the
pre-exponential factor $J$ in the main order in $T_{L,R}$:

\begin{equation}\label{AA1}
$$J\approx\exp\left(\sum_{i=L,R}{1\over 4}\ln T_i+const\right)\approx
C\left(T_LT_R\right)^{1/4}.$$
\end{equation}

\section{Appendix B: Pre-exponential factor for 2D case}

In the 2D case, the spectrum $\{\chi (n,m)\}$ of fluctuations around the
saddle-point should be classified by orbital and radial quantum
numbers $n$ and $m$, respectively, and can be found from the
eigenvalues of the Hamiltonian
$$\hat {H}=-r^{-1}\partial_r(r\partial_r)+m^2r^{-2}+U,$$
where $U$ is determined by Eq. (\ref{IIID7}).

Without any symmetry-breaking, the spectrum of $\chi$'s can be
approximated as

\begin{equation}\label{AB1}
$$\chi (0,0)\approx 2L^{-2}/\ln (L/l)$$
\end{equation}
for the lowest mode and as

\begin{equation}\label{AB2}
$$\chi (n,m)\approx\left({{\pi}\over L}\right)^2\left[n+{1\over 4}
+{m\over 2}\right]^2$$
\end{equation}
for higher $n$'s and $m$'s.

The optimal solution breaks the fermion-boson symmetry and induces
effective potentials composed of two types of contributions,
$${1\over 4}k(\partial\theta_t)^2\approx k\mu^2/r^2\;{\rm a}{\rm n}
{\rm d}\quad\kappa{t\over {2\pi\nu D}}e^{-\theta_t}\approx\kappa\mu
L^{-2}({L\over r})^{2\mu}.$$
In the above equations, the approximate values are given for the
most important range of distances $r\le L\sqrt {z(T)/T}$, and one has to
remember that $\mu <1$.

For any $m\ne 0$, the potential $U$ is relatively small, $U\ll m^2
/r^2$, and
could be treated perturbatively.  Due to the sum rule mentioned in
Section III.D, Eq.  (\ref{IIID8}), the modes with $m\ne 0$ contribute only
in the second order in $U$, and what they give to the exponential of $
J$
is of the order of $\mu^4\ln (L/l);\mu^2$.  With the accuracy we need here
regarding the leading terms in $F_t$, this contribution can be neglected.

The spectrum of low-lying modes $\{\chi (n>0,0)\}$ is given by the
expression

\begin{equation}\label{AB3}
$$\chi (n,0)\sim\left({{\pi}\over L}\right)^2\left[n+{1\over 4}+\sqrt
k{{\mu}\over 2}\right]^2,\quad 0<n\le{L\over {\pi l}}.$$
\end{equation}

The cancellation between different eigenvalues from Eqs.
(\ref{AB1}-\ref{AB3}) substituted to the general equation in Eq.
(\ref{IIID5}) produces a multiplier to $J$ which is of the order of $
e^{\sim\mu^2}$
for $T\ll 1$ and $\mu\ln{L\over l}$ at $T\gg 1$.  Finally, we get
$$\matrix{J=1+0(T^2),\;T\ll 1\cr
J\propto\mu\exp (\mu\ln{L\over l})\sim T,\;T\gg 1\cr}
.$$
This result can be used for all symmetry classes.


\begin{references}
\bibitem{Rem1}
'Intrinsic' conductance $g$ depends on the dimensionality of the system:
$$1{\rm D}:\;g\sim{{L_c}\over L};\quad 2{\rm D}:\;g\sim{{p_Fl}\over {\ln
(L/l)}};\quad 3{\rm D}:\;g\sim (p_Fl)^2.$$
Here, we speak about the dimensionless conductance, $g\equiv G/(e^
2/h)$.
\bibitem{Rem2}
It is worth mentioning that $f(\downarrow ,t)$ is normalized, $t_0
=1$, and the
normalization of spin-projected wave functions, $t_1=1/(2V)$, produces
the same requirement as in Eq.  (\ref{IIIA5})
\bibitem{Spin-unitary}
By this term, we denote the class of the systems both with the
spin-dependent mechanism of scattering and broken time-reversal
symmetry.  For instance, this can be imagined as a metal with
built-in disordered static magnetic field or a random exchange field.
\end{references}
\end{document}